\documentclass{epl}
\usepackage{epsfig}
\usepackage{graphicx}

\title{Towards a continuum theory of clustering in a freely cooling inelastic gas}
\shorttitle{Clustering in an inelastic gas}
\author{Baruch Meerson\inst{1} \and Andrea Puglisi\inst{2}}

\institute{\inst{1} Racah Institute of Physics, Hebrew University
of Jerusalem, Jerusalem 91904, Israel \\
\inst{2} Laboratoire de Physique Th\'{e}orique (UMR 8627 du CNRS),
B\^{a}timent 210, Universit\'{e} de Paris-Sud, 91405 Orsay Cedex,
France}



\begin{document}

\maketitle

\begin{abstract}
We performed molecular dynamics simulations to investigate the
clustering instability of a freely cooling dilute gas of
inelastically colliding disks in a quasi-one-dimensional setting.
We observe that, as the gas cools, the shear stress becomes
negligibly small, and the gas flows by inertia only. Finite-time
singularities, intrinsic in such a flow, are arrested only when
close-packed clusters are formed. We observe that the late-time
dynamics of this system are describable by the Burgers equation
with vanishing viscosity, and predict the long-time coarsening
behavior.
\end{abstract}

\section{Introduction}

A gas of hard spheres is a standard model of statistical physics
and kinetic theory \cite{LL}. It is surprising that a minor change
in this model - the introduction of energy loss in the binary
collisions - leads to consequences so dramatic. Among the many
fascinating properties of the gas of \textit{inelastically}
colliding hard spheres \cite{granulargas}, the clustering
instability \cite{Goldhirsch2,McNamara} plays a special role. No
matter how small (but finite) is the inelasticity of collisions,
the homogeneous cooling state (HCS) of this gas is always unstable
if the system size $L$ is large enough. When studying macroscopic
properties of matter, a physicist deals, first of all, with the
thermodynamic limit $L \to \infty$. From this perspective the gas
of \textit{elastically} colliding hard spheres is a singular limit
of the inelastic gas problem. The practical importance of the
inelastic gas model stems from its being the simplest model of
granular flow \cite{granulargas,Haff,Campbell}.

The clustering instability of a freely cooling inelastic gas
involves formation of clusters of particles and generation of
vortices \cite{Goldhirsch2,McNamara,Barrat}. The basic physics of
the initial stage of cluster formation is simple: the inelastic
cooling of the gas causes a pressure drop in the regions of
enhanced density. This pressure drop drives an inflow of gas from
the periphery and therefore provides a positive feedback to the
instability. A traditional framework for quantitative theory here
is \textit{granular hydrodynamics} (GH), which assumes scale
separation and is derivable from the Boltzmann equation, properly
modified to account for the inelasticity of collisions
\cite{granulargas}. Though the general criteria of its validity
remain controversial \cite{Kadanoff}, GH is well established at
least when the following two criteria are met: (i) the granular
gas is dilute, $n \sigma^D \ll 1$, and (ii) the particle
collisions are nearly elastic, $q \ll 1$. Here $n$ is the local
number density of the gas, $\sigma$ is the particle diameter,
$D>1$ is the dimension of space, $q=(1-r)/2$ is the inelasticity
of collisions, and $r$ is the coefficient of normal restitution
(assumed constant throughout this paper). In this case the GH
equations, linearized around the HCS, provide an accurate theory
of the initial stage of the instability in terms of the (linear)
hydrodynamic modes of the system: the shear mode (spontaneous
formation of vortices) and the entropy, or clustering mode
(formation of clusters) \cite{Goldhirsch2,McNamara,Barrat}.

The further evolution of the instability is a hard problem. One
difficulty here is technical, as the growing nonlinear shear and
clustering modes become strongly coupled. Another difficulty is
conceptual: GH breaks down in high-density regions. All previous
attempts to develop a theory beyond linearization of the GH
equations attempted to circumvent these difficulties. Ben-Naim
\textit{et al.} \cite{Ben-Naim1} considered \textit{point-like}
particles inelastically colliding on a \textit{line}. This
strictly one-dimensional (1D) geometry makes a GH description
problematic \cite{Kadanoff}. Still, Ben-Naim \textit{et al.}
observed that, at long times, the 1D system is describable by the
Burgers equation with vanishing viscosity.  Ernst \textit{et al.}
\cite{ErnstTDGLE} considered a small two-dimensional (2D) system,
where the entropy mode is suppressed, and dealt with the unstable
shear mode. Baldassarri \textit{et al.}~\cite{Baldassarri} also
studied instability of the velocity field in a homogeneous gas, by
introducing a lattice model.

Focusing on the entropy, or clustering, mode, Efrati \textit{et
al.} \cite{ELM} had the shear mode  suppressed, by working with a
\textit{quasi}-1D setting, see below. They solved numerically the
low-density GH equations and found that, as the unstable system
cools down, the shear stress becomes negligibly small, and a
\textit{flow by inertia} sets in. Formally, this flow develops a
finite-time singularity: the velocity gradient and the gas density
diverge at some location. Efrati \textit{et al.} argued that the
flow by inertia is an important intermediate stage
of the clustering instability. However, 
they did not address still later stages of the instability, when
finite-density effects come into play. We report here the first MD
simulations which probe the quasi-1D clustering of a dilute
inelastic gas. We find that the simulated low-density stage of the
clustering instability is in excellent agreement with the
hydrodynamic predictions of Ref. \cite{ELM}. What happens at later
times? We observe that the attempted singularities are arrested
only when hexagonally packed clusters are formed. The still later
dynamics are describable by the Burgers equation with vanishing
viscosity, in a striking analogy with the purely 1D result by
Ben-Naim \textit{et al.} \cite{Ben-Naim1}. Based on this
observation, we predict the very-late-time coarsening dynamics of
the system, all the way to its simple final steady state.

\section{Model system and clustering instability}

Consider an assembly of $N$ identical hard disks of mass $m=1$,
diameter $\sigma =1$, and inelasticity $0<q \ll 1$ in a 2D box
with dimensions $L_x$ and $L_y$ ($L_x \gg L_y$). The initial
number density of the gas $n_0=N/(L_x L_y)$ is very small compared
to the hexagonal close packing density $n_c = 2/(\sqrt{3}
\sigma^2) \simeq 1.155$. The initial particle velocity
distribution is gaussian with temperature $T_0=1$. The parameters
are chosen so that the low-density GH equations \cite{Haff} are
accurate until relatively late times, when the local value of $n
\sigma^2$ is not a small parameter anymore. The boundary
conditions are periodic at $x=0$ and $x=L_x$ and elastic (specular
reflection) at $y=0$ and $y=L_y$. Assuming a homogeneous cooling,
one arrives at Haff's law \cite{Haff} $T(t)=T_0(1+t/t_0)^{-2}$,
where $t_0=(2 \pi^{1/2} \sigma n_0 q T_0^{1/2})^{-1}$ is the
characteristic cooling time. The GH equations, linearized around
the HCS, show that the HCS is unstable if $L_x$ and/or $L_y$ are
large enough \cite{Goldhirsch2,McNamara,Barrat}. The density
perturbations grow in time algebraically. The temperature and
velocity perturbations decay, but the decay rates are smaller than
that described by Haff's law. As a result, the flow tends to
become supersonic \cite{Barrat}. A strong quasi-1D instability
requires two criteria: $\sigma n_0 L_x q^{1/2}\gg 2 \pi^{1/2}$,
and $\sigma n_0 L_y q^{1/2}\ll \pi^{1/2}$; the latter one
guarantees suppression of the shear and clustering modes in the
$y$-direction \cite{Goldhirsch2,McNamara,Barrat,ELM}.

\section{Flow by inertia}

\begin{figure}[htbp]
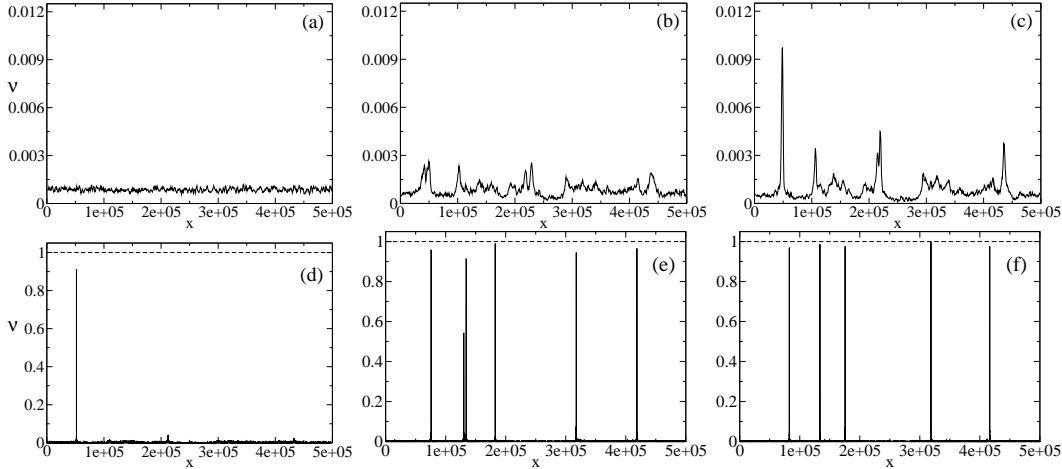

\begin{center}
\includegraphics[width=4.6cm,clip=true]{n_1.eps}
\includegraphics[width=4.6cm,clip=true]{n_50.eps}
\includegraphics[width=4.6cm,clip=true]{n_60.eps}\\
\includegraphics[width=4.6cm,clip=true]{n_100.eps}
\includegraphics[width=4.6cm,clip=true]{n_1000.eps}
\includegraphics[width=4.6cm,clip=true]{n_2400.eps}
\end{center}
\caption{The local area fraction $\nu(x,t) = n(x,t)/n_c$ at times
$0$ (a), $274\,622$ (b), $549\,554$ (c), $771\,055$ (d), $2.14972
\times 10^6$ (e), and $2.53252 \times 10^6$ (f).}
\label{fig:density}
\end{figure}

\begin{figure}[htbp]
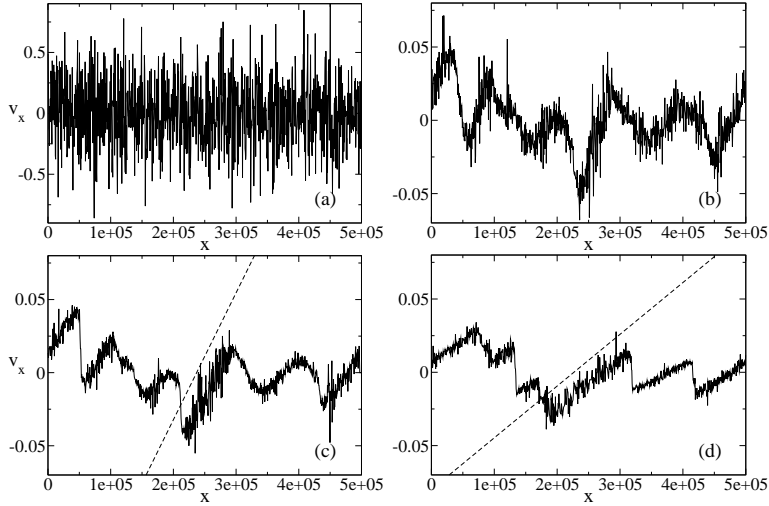

\begin{center}
\includegraphics[width=5cm,clip=true]{vx_1.eps}
\includegraphics[width=5cm,clip=true]{vx_50.eps}\\
\includegraphics[width=5cm,clip=true]{vx_100.eps}
\includegraphics[width=5cm,clip=true]{vx_2400.eps}
\end{center}
\caption{The horizontal velocity profiles at times $0$ (a),
$274\,622$ (b), $ 771\,055$ (c), and $2.53252 \times 10^6$ (d).
The straight lines in Figs. c and d have a slope $1/(t+C)$, where
$C=3.8 \times 10^5$.} \label{fig:velocity}
\end{figure}

We performed extensive event-driven MD simulations in this regime
of parameters. We verified that no structure in the $y$-direction
appears, as expected. Therefore, our diagnostics focused on 1D
coarse-grained fields: the density $n(x,t)$, the mean velocity
$\mathbf{v}(x,t)=\left(v_x, v_y\right)$ and the $x$- and $y$-
components of the velocity fluctuations: $T_x(x,t)$ and
$T_y(x,t)$. We report here a typical simulation with $N=12\,500$,
$q=0.04$, $L_x =5 \times 10^5$ and $L_y = 25$. The initial gas
density is $n_0 = 10^{-3}$, while $t_0 \simeq 7.05 \times 10^4$.
Strong clustering instability is clearly seen in Fig.
\ref{fig:density}, which exhibits formation of multiple clusters,
and Fig. \ref{fig:velocity}, which shows an inflow of gas into the
forming clusters. Meanwhile, the gas temperature $T=T_x+T_y$ (not
shown) rapidly decays with time. Throughout the simulated dynamics
$T_x$ and $T_y$ remain close to each other, while their spatial
inhomogeneity is small compared to the strong inhomogeneity of
$v_x$ and $n$.

The average particle energy versus time, $E_{tot}(t)$, follows
Haff's law at early times, but deviates from it at later times,
when strong hydrodynamic motions develop, see Fig.
{\ref{fig:energies} (left). In the hydrodynamic description
$E_{tot}(t) = (n_0 L_x)^{-1} \int_0^{L_x} (E_{T}+E_{mac}) \,dx$,
where $E_{T}=nT$ is the thermal energy density, and $E_{mac}=n
v^2/2 \equiv n (v_x^2+v_y^2)/2$ is the macroscopic kinetic energy
density.  The role of each term is elucidated in Fig.
{\ref{fig:energies}. Both $E_{T}(t)$, and $E_{mac}(t)$ initially
decay; $E_{T}(t)$ decays faster. At later times $E_{T}(t)$ and the
$y$-component of $E_{mac}(t)$ continue to decay rapidly, while the
$x$-component of $E_{mac}(t)$ decays much slower. As a result,
$E_{tot}(t)$ is dominated by $E_{T}$ at early times and by the
$x$-component of $E_{mac}(t)$ at later times. Remarkably,
$E_{T}(t)$ continues to follow Haff's law quite closely until the
latest simulated times.



\begin{figure}[htbp]
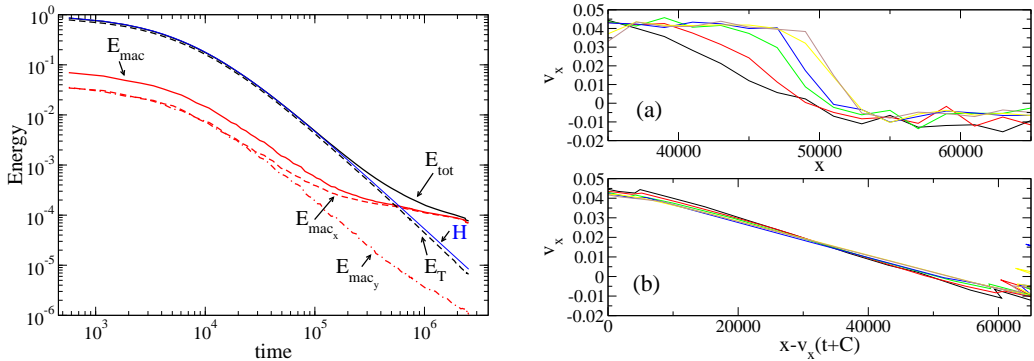

\begin{center}
\parbox{6.5cm}{\includegraphics[width=6.5cm,clip=true]{energies.eps}}
\hspace{0.5cm}
\parbox{6.5cm}{\includegraphics[width=6.5cm,clip=true]{collapseB.eps}}
\end{center}
\caption{{\bf Left}: the energy balance of the system versus time.
  $E_{tot}$ is the average kinetic energy of the particles, $E_T$ is the
  average thermal energy, and $E_{mac}$ is the average energy of the
  macroscopic motions. The Haff's law is indicated by $H$. {\bf Right}: The
  horizontal velocity profiles in a region around the leftmost peak. Shown are
  $v_x$ versus $x$ (a), and $v_x$ versus $x-(t+C)\,v_x(x,t)$(b) at times
  $294518$, $410692$, $549554$, $636990$, $658371$, $680869$ and $701507$ with
  $C=3.8 \times 10^5$.} \label{fig:energies}
\end{figure}

\begin{figure}[htbp]
\begin{center}
\includegraphics[width=10cm,clip=true]{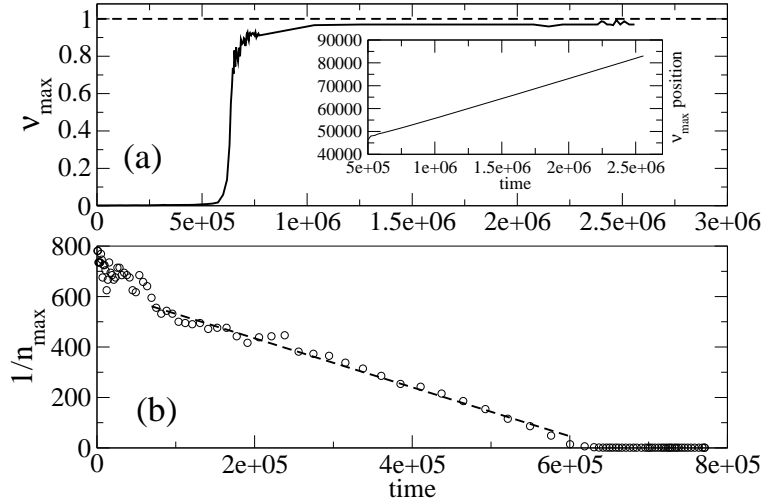}
\end{center}
\caption{Evolution of the leftmost peak in
Fig.~\ref{fig:density}f. Figure a shows $\nu_{max}$ versus time.
Figure b shows $n_{max}^{-1}$ versus time and a linear fit. The
inset shows the $x$-coordinate of the peak versus time.}
\label{fig:maximum}
\end{figure}

Figure {\ref{fig:maximum}a shows the time history of a typical
cluster (the leftmost density peak in Fig. {\ref{fig:density}
c-f). The rapid density growth is saturated when $n_{max}$
approaches $n_c$.  A snapshot of the density peak region (Fig.
{\ref{fig:snap}) indeed shows almost perfect hexagonal packing.
The rapid density growth is shown in more detail in Fig.
{\ref{fig:maximum}b which depicts $1/n_{max}$ versus time. The
linear dependence, observed at intermediate times,  indicates an
``attempted" finite-time singularity $n_{max} \sim (const -
t)^{-1}$. The same density singularity was observed in
hydrodynamic simulations \cite{ELM}; it is caused by a flow by
inertia which develops when the forces acting on a fluid element
vanish. The reason for it in the freely evolving inelastic gas is
the continued rapid cooling, which makes the pressure and viscous
stresses negligible \cite{ELM}. The flow-by-inertia equations read
\begin{equation} \label{inertia}
\frac{\partial v_x}{\partial t} + v_x\,\frac{\partial
v_x}{\partial x} = 0\,,\,\,\, \mbox{(a)}
\;\;\;\;\;\mbox{and}\;\;\;\;\; \frac{\partial n}{\partial t} +
\frac{\partial (n v_x)}{\partial x} = 0\,. \,\,\, \mbox{(b)}
\end{equation}
These equations are soluble analytically in Lagrangian coordinates
\cite{Whitham,Zeldovich1}:
\begin{equation}\label{solution}
v_x(x,t) = v_0(\xi)\,,\,\,\,\,\mbox{(a)} \;\;\;\;\;\;
n(x,t)=\frac{n_0(\xi)}{1+(t+C)\,v_0^{\prime}(\xi)}\,.
\,\,\,\,\,\mbox{(b)}
\end{equation}
where $C$ is an arbitrary constant, $v_0^{\prime}(\xi)\equiv d
v_0(\xi)/d \xi$, while $v_0(\xi)$ and $n_0(\xi)$ are the $v_x$ and
$n$, respectively, at the ``initial" moment of time $t=-C$. The
relation between Eulerian coordinate $x$ and Lagrangian coordinate
$\xi$ is $x=\xi+ v_0(\xi)\,(t+C)$. The finite-time singularities
of both the velocity gradient, and the density occur when the
denominator in Eq. (\ref{solution}) b vanishes for the first time;
it requires $v_0^{\prime}(\xi)<0$.

The results of our MD simulations quantitatively agree with the
flow-by-inertia scenario. Using in Eq. (\ref{solution}b) the
straight-line fit, shown in Fig. {\ref{fig:maximum}b, we obtain $C
\simeq 3.8 \times 10^5$ and $|v_0(\xi_*)|  \simeq 1.0 \times
10^{-6}$, where $\xi_*$ is Lagrangian coordinate of the density
peak. Now we verify Eq. (\ref{solution}) by observing [see Fig.
{\ref{fig:energies} (right)] collapse of $v_x (x,t)$ versus
$\xi=x-(t+C)\,v_x(x,t)$, at different times, in the region of the
leftmost density peak. The slope of the obtained straight line
yields an independent estimate of $|v_0(\xi_*)|$ which agrees
within 3 percent with the value of $1.0 \times 10^{-6}$ found
earlier. Still another prediction deals with the shape of the
density peak close to the time of attempted singularity. The
flow-by-inertia theory predicts a density profile $n \sim
|x-x_{max}|^{-2/3}$ \cite{Zeldovich1}. The simulations confirm
this prediction, see Fig. {\ref{fig:snap} (right).



\begin{figure}[htbp]
\begin{center}
\parbox{6.0cm}{\includegraphics[width=6.5cm,clip=true]{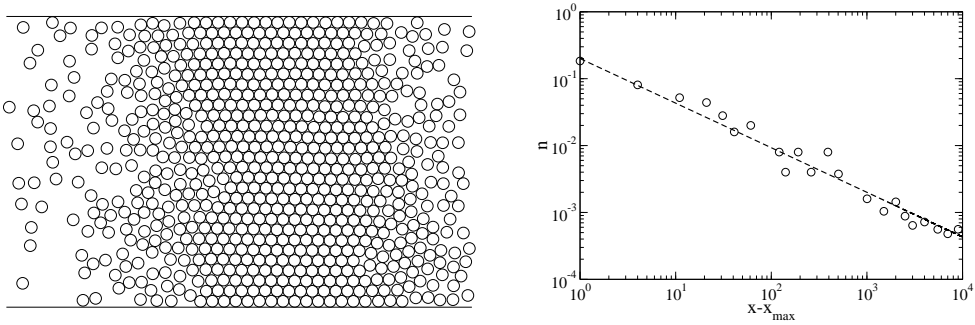}}
\hspace{1cm}
\parbox{6.0cm}{\includegraphics[width=5.8cm,clip=true]{tail.eps}}
\end{center}
\caption{\textbf{Left:} a snapshot of the system in the region of
the leftmost peak at time $771\,055$. \textbf{Right:} The right
tail of the density peak close to the time of attempted
singularity. Shown is the log-log plot of the density profile $n$
of the leftmost density peak versus $x-x_{max}$ at time
$732\,951$. The dashed line is a power law $0.2 \times
(x-x_{max})^{-2/3}$.} \label{fig:snap}
\end{figure}

As we observed, the density growth in the clusters is suppressed
only when the density approaches $n_c$. At a fixed temperature,
the pressure of an assembly of nearly elastic hard spheres
diverges like $(n_c-n)^{-1}$ as $n$ approaches $n_c$. Is there a
pressure ``revival" at $n\to n_c$? We used the empiric relation
$p= nT (n_c+n)/(n_c-n)$, which interpolates between the dilute
limit and the close-packing limit \cite{Grossman}, to calculate
the pressure field $p(x,t)$. We found that, though $n_{max}$
approaches $n_c$,  $p(x,t)$ continues to decay with time,
apparently because of the very rapid temperature decay. Therefore,
the pressure revival mechanism can be ruled out.

\section{The Burgers model}
As the system approaches close packing it gets jammed. Faster
particles cannot overrun slower ones, and the singular growth of
the velocity gradient and density is arrested. A natural continuum
model for a jammed flow is the Burgers equation \cite{Whitham}
\begin{equation} \label{burgers}
\frac{\partial v_x}{\partial t} + v_x\,\frac{\partial
v_x}{\partial x} = \nu \frac{\partial^2 v_x}{\partial x^2}
\end{equation}
in the limit of vanishing viscosity $\nu \to 0$, together with the
continuity equation (\ref{inertia})b. We shall call this model
``the Burgers model". Equation (\ref{burgers}) is soluble exactly
by the Hopf-Cole transformation \cite{Burgers,Gurbatov}. The
zero-viscosity limit of the solution, which is the subject of our
interest, has the following properties. The solution is
\textit{identical} to that predicted by the flow-by-inertia model
until the time moment when the flow by inertia would have
developed a singularity. The attempted singularity gives way, in
the Burgers model, to a ``shock": a jump in the velocity field
which carries a density peak (cluster). In this coarse-grained
description, the close-packed clusters have zero sizes but finite
masses. At long times, when the shocks have ``matured"
\cite{Gurbatov}, the quantitative predictions of the Burgers model
are especially simple and can be conveniently tested in our MD
simulations. One prediction is that $\partial v_x/\partial x$ is
equal to $1/(t+C)$ everywhere between the shocks, where $C$ is the
same constant ($\simeq 3.8 \times 10^5$ in the reported
simulation) as above. This prediction is tested in Fig.
\ref{fig:velocity}c and d. While the agreement in Fig. c is only
fair (as the shocks have not yet matured), it improves
considerably in Fig. d. Another prediction that we verified (see
the inset of Fig. \ref{fig:maximum}a) is that each shock moves
with a constant speed until it collides with another shock.
Furthermore, a collision between two shocks leads to their merger,
without any thermalization of the system. Such a merger event can
be seen  in the left part of Figs. \ref{fig:density} e and f.
Finally, we verified that the zeros of $v_x$, belonging to the
``ramps", stay at rest as expected \cite{Gurbatov}.

\section{Late-time coarsening dynamics}
Event-driven MD simulations become very slow once the hexagonal
close packing in the clusters is achieved. Fortunately, the times
reached in our simulations were large enough to have verified the
Burgers model as a proper late-time continuum model of the
quasi-1D clustering process. Therefore, we can give a detailed
prediction of the still later \textit{coarsening} dynamics of the
system, without a need to simulate it. If the number of clusters
is large, the coarsening dynamics, which proceed via cluster
mergers, can be addressed statistically. At this level of
description the problem coincides with that of the decaying
Burgers turbulence \cite{Burgers,Gurbatov}, or the ballistic
agglomeration model \cite{Carnevale}. For uncorrelated initial
conditions, the average cluster mass grows with time like
$t^{2/3}$, the average velocity decreases like $t^{-1/3}$, and the
average distance between two neighboring clusters grows like
$t^{2/3}$ \cite{Carnevale}. This yields a long-time asymptotic
scaling law $E_{tot} \simeq E_{v_x} \sim t^{-2/3}$. At this stage,
$E_{tot}$ is dominated by the kinetic energy of the clusters. The
energy decay stops when an ultimate steady state of the system is
reached: a single ``super-cluster", moving with a small constant
speed (which can be determined from the initial data, by employing
the momentum conservation in the $x$-direction and assuming that
\textit{all} $N$ particles are absorbed by this super-cluster).

\section{Summary and Discussion}
Our MD simulations fully support the hydrodynamic flow-by-inertia
scenario \cite{ELM}. The attempted singularities of this flow are
suppressed when almost perfect hexagonally packed clusters are
formed. At still later times the dynamics are describable by the
Burgers equation with vanishing viscosity, in a striking analogy
with the purely 1D results by Ben-Naim \textit{et al.}
\cite{Ben-Naim1}. Therefore, clustering  instability in a quasi-1D
setting is now well understood. What about a fully
multi-dimensional geometry? It has been conjectured that, prior to
the first attempted singularity, the system should be describable
by a \textit{multi-dimensional} flow-by-inertia model \cite{ELM}.
One can expect that an interplay between vorticity, produced
during the early stage of the instability, and jamming provides a
saturation mechanism for the multi-dimensional singularities.
Unfortunately, the multi-dimensional Burgers model
\cite{Burgers,Gurbatov} assumes a potential velocity field and
misses the important physics, caused by the presence of vorticity.
The formulation of a continuum model free of this flaw should be
the next step of theory.

\acknowledgments

We thank Efi Efrati for a useful discussion. The work was
supported by grants from the Israel Science Foundation and
German-Israeli Foundation for Scientific Research \& Development,
and by the Marie Curie grant No. MEIF-CT-2003-500944.


\begin{thebibliography}{0}
\bibitem{LL} \Name{Landau L.D. \and Lifshitz E.M.} \Book{Statistical Physics, Part 1}
\Publ{Pergamon, Oxford} \Year{1980}; \Name{Lifshitz E.M. \and
Pitaevskii L.P.} \Book{Physical Kinetics} \Publ{Pergamon, Oxford}
\Year{1981}.
\bibitem{granulargas} \Book{Granular Gases}\Editor{P\"{o}schel T. \and Luding S.}
\Publ{Springer, Berlin} \Year{2001}; \Book{Granular Gas Dynamics}
\Editor{P\"{o}schel T. \and Brilliantov N.} \Publ{Springer,
Berlin} \Year{2003}; \Name{Goldhirsch I.} \REVIEW{Annu. Rev. Fluid
Mech.} {35} {2003} {267}.
\bibitem{Goldhirsch2} \Name{Goldhirsch I. \and
Zanetti G.} \REVIEW{Phys. Rev. Lett.} {70} {1993} {1619}.
\bibitem{McNamara} \Name{McNamara S.} \REVIEW{Phys. Fluids A} {5} {3056}; \Name{McNamara S. \and Young W. R.} \REVIEW{Phys. Rev.
E} {53} {1996} {5089}.
\bibitem{Haff} \Name{Haff P.K.} \REVIEW{J. Fluid Mech.} {134} {1983} {401}.
\bibitem{Campbell} \Name{Campbell C.S.} \REVIEW{Annu. Rev. Fluid Mech.} {22} {1990} {57}.
\bibitem{Barrat} \Name{Deltour P. \and  Barrat J.-L.} \REVIEW{J. Phys I France} {7} {1997} {137}.
\bibitem{Kadanoff} \Name{Kadanoff L.P.} \REVIEW{Rev. Mod. Phys.} {71} {1999} {435}.
\bibitem{Ben-Naim1} \Name{Ben-Naim E., Chen S.Y., Doolen G.D. \and
Redner S.} \REVIEW{Phys. Rev. Lett.} {83} {1999} {4069}.
\bibitem{ErnstTDGLE} \Name{Wakou J., Brito R., \and Ernst M.H.} \REVIEW{J. Stat. Phys.} {107} {2002} {3}.
\bibitem{Baldassarri} \Name{Baldassarri A., Bettolo U., \and Puglisi A.}
  \REVIEW{Europhys. Lett.} {58} {2002} {14};
  \REVIEW{Phys. Rev. E} {65} {2002} {051301}.
\bibitem{ELM} \Name{Efrati E., Livne E., \and Meerson B.} \REVIEW{Phys. Rev. Lett} {94} {2005} {088001}.
\bibitem{Whitham} \Name{Whitham G.B.} \Book{Linear and Nonlinear Waves}
\Publ{Wiley, New York} \Year{1974}.
\bibitem{Zeldovich1} \Name{Shandarin S.F. \and Zeldovich Ya. B.} \REVIEW{Rev. Mod. Phys.} {61} {1989} {185}.
\bibitem{Grossman} \Name{Grossman E. L., Zhou T., \and Ben-Naim E.} \REVIEW{Phys. Rev. E} {55} {1997} {4200}.
\bibitem{Burgers} \Name{Burgers J.M.} \Book{The Nonlinear Diffusion
Equation} \Publ{Reidel, Dordrecht} \Year{1974}.
\bibitem{Gurbatov} \Name{Gurbatov S.N., Malakhov A.N., \and Saichev A.I.} \Book{Nonlinear
Random Waves and Turbulence in Nondispersive Media}
\Publ{Manchester University Press, Manchester} \Year{1991}.
\bibitem{Carnevale} \Name{Carnevale G.F., Pomeau Y. \and Young
W.R.} \REVIEW{Phys. Rev. Lett.} {64} {1990} {2913}.
\end{thebibliography}
\end{document}